\documentclass[namedreferences,hyperref,optionalrh,solaromanenum]{spr-sola}

\usepackage{graphicx}                    
\usepackage{amssymb}                    
\usepackage{amsmath}
\usepackage{color}                       
\usepackage{breakurl}                         

\newcommand{\I}{\mathcal{I}}
\newcommand{\V}{\mathcal{V}}

\newcommand{\ml}{m-$\lambda$}
\newcommand{\dml}{dm-$\lambda$}
\newcommand{\cml}{cm-$\lambda$}

\newcommand{\rms}{{\sl rms}\rm}
\chardef\us=`\_

\begin{document}
\begin{article}
\begin{opening}

\title{Noise in Maps of the Sun at Radio Wavelengths \\ II: Solar Use Cases}

%
\author[addressref={aff1},corref,email={tbastian@nrao.edu}]{\inits{T. S. }\fnm{Timothy }\snm{Bastian}\orcid{0000-0002-0713-0604}}
\author[addressref={aff2},corref,email={bin.chen@njit.edu}]{\inits{B. }\fnm{Bin }\snm{Chen}\orcid{}}
\author[addressref={aff2},corref,email={surajit.mondal@njit.edu}]{\inits{S. }\fnm{Surajit }\snm{Mondal}\orcid{0000-0002-2325-5298}}
\author[addressref={aff3},corref,email={shilaire@berkeley.edu}]{\inits{P. }\fnm{Pascal }\snm{Saint-Hilaire}\orcid{}}

%
\runningauthor{Bastian et al.}
\runningtitle{Noise in Radio Maps of the Sun \\ II: Science Use Cases}

\address[id={aff1}]{National Radio Astronomy Observatory, 520 Edgemont Rd, Charlottesville, VA 22903, USA}
\address[id={aff2}]{Center for Solar-Terrestrial Research, New Jersey Institute of Technology, 323 M L King Jr Boulevard, Newark, NJ 07102, USA}
\address[id={aff3}]{Space Sciences Laboratory, University of California, 7 Gauss Way, Berkeley, CA, 94720 USA}

\begin{abstract}
Noise in images of strong celestial sources at radio wavelengths using Fourier synthesis arrays can be dominated by the source itself, so-called self-noise. We outlined the theory of self-noise for strong sources in a companion paper. Here we consider the case of noise in maps of radio emission from the Sun which, as we show, is always dominated by self noise. We consider several classes of science use cases for current and planned arrays designed to observe the Sun in order to understand limitations imposed by self-noise. We focus on instruments operating at decimeter and centimeter wavelengths but the results are applicable to other wavelength regimes. 
\end{abstract}


\end{opening}


\pagebreak

 \section{Introduction}\label{Intro} 
 
Modern imaging observations of the Sun at radio wavelengths exploit interferometric arrays designed to perform Fourier synthesis mapping. The Sun represents an extremely challenging imaging problem \citep{Bastian1989}. It emits on angular scales ranging from less than an arcsecond to more than half a degree; its emissions vary on a wide range of time scales, from $\sim\!10$ ms to many years; it produces spectroscopically rich emission with complex and dynamic structure; and it produces highly polarized emission. A key limitation to high-fidelity, high-dynamic range imaging of the Sun using radio Fourier synthesis techniques has been the number and distribution of antennas in arrays currently in use for solar imaging, especially at decimeter (\dml) and centimeter (\cml)  wavelengths, the focus of our discussion here. With a limited number $n$ of antennas, or a non-optimal distribution of antennas, the Sun's complex brightness distribution is under-sampled or incompletely sampled, resulting in missing spatial information, poor image reconstruction, and high levels of residual sidelobes. 


Next-generation radio imaging instruments capable of observing the Sun must be designed and optimized to address solar imaging requirements which are, in turn, based on science requirements. As we discuss in later sections, high signal-to-noise ratios are needed to achieve many science objectives. It is therefore important to consider all instrumental factors that may limit performance. These include the design of the constituent antennas in an imaging array, their number and distribution, the system electronics and their stability, instrumental calibration, and noise. The latter includes both instrumental noise and, in the case of the Sun, noise due to the source itself, so-called ``self-noise". For radio imaging arrays designed to observe faint celestial sources, great pains are taken to ensure that instrumental noise is minimized. Since self-noise is caused by the source itself, the limitations imposed by such noise on imaging performance for a given synthesis imaging array must also be understood.  

In this paper we focus on noise in Fourier synthesis images of the Sun and consider the limitations imposed by self-noise on an otherwise ideal system. We introduce Fourier synthesis imaging at radio wavelengths and outline the general theory of self-noise in a companion paper (Paper I). In this paper we summarize key results from Paper I in Section~2 so that the current paper is largely self-contained. We briefly describe several Fourier synthesis arrays that are currently in use for solar observations, or are in the planning stages, in Section~3.  We then show in Section~4 that that self-noise must be considered for all current and planned arrays designed to observe the Sun at the wavelengths in question.  In Section~5 we consider various science use cases in the so-called snapshot imaging regime for two Fourier synthesis arrays currently operating at radio wavelengths.  We summarize our results in Section~6 and conclude with a brief assessment of the impact on self-noise of next-generation radio telescopes capable of observing the Sun. 

\section{Self-noise in Fourier Synthesis Maps}

The theory of self noise in Fourier synthesis maps is outlined in Paper~I. While the Sun has been observed at radio wavelengths for decades, self-noise has not received much attention as a limiting factor because other systematic effects are typically more significant. These include, for example, calibration errors and/or deconvolution errors due to incomplete sampling of the aperture plane. Nevertheless, self-noise represents a fundamental limit to imaging performance and is our focus here. 

We denote instrumental noise associated with a given antenna by $N$ and the total source flux density by $S$. The former is the same as the {\sl source equivalent flux density} and is related to the {\sl system temperature} $T_{\rm sys}$ through $N=T_{\rm sys}/K$, where $K=A_e/2k_B$; $A_e$ is the antenna effective area and $k_B$ is Boltzmann's constant. The spectral flux density $S$ incident on an antenna contributes incremental noise to the system characterized by the {\sl antenna temperature} $T_{\rm ant}=SA_e/2k_B$ and so we have $S=T_{\rm ant}/K$. Further details can be found in Paper~I. $S$ and $N$ are typically measured in units of Jansky\footnote{1~Jy $\equiv\ 10^{-26}$ ergs cm$^{-2}$ s$^{-1}$ ster$^{-1}$} or solar flux units (1~SFU $\equiv\ 10^4$~Jy) while $T_{\rm sys}$ and $T_{\rm ant}$ are expressed in Kelvin. 

A correlation array with $n$ antennas that includes total power measurements (which we referred to as a total power array in Paper I) produces a dirty map can be expressed in terms of a direct Fourier transform:

\begin{eqnarray}
\I_D^\circ(\theta_x,\theta_y)=\frac{1}{n^2}\bigl\lbrace \sum_{i=1}^n Z_i &+& \sum_{j=1}^{n} \sum_{k>j}^{n}[\V_{jk}(u,v)e^{-i(\theta_x u+\theta_y v)}\\ \nonumber
&+&\V_{jk}^\ast(u,v)e^{+i(\theta_x u+\theta_y v)}]\bigr\rbrace,
\end{eqnarray}

\noindent where $\V_{jk}(u,v)$ is the complex visibility measured by an antenna pair, or antenna baseline, $jk$; $u=\nu(x_j-x_k)/c$, and $v=\nu(y_j-y_k)/c$ are the antenna baseline coordinates in the aperture domain; $\nu$ is the observing frequency; and $(x_j,y_j)$ is the spatial coordinate of antenna $j$. Each antenna in a total power array also measures $Z_i=S_i+N_i$ the signal and instrumental noise measured by antenna $i$ which, for identical antennas, we take to be $S$ and $N$. The map \rms\ can then be expressed simply as 

\begin{equation}
\sigma_I(\theta_x,\theta_y) = \frac{1}{M}\biggl[\I_D^\circ(\theta_x,\theta_y)+\frac{N}{n}\biggr],
\end{equation}

\noindent where $M=\sqrt{\Delta\nu\tau}$, $\Delta\nu$ is the frequency bandwidth, and $\tau$ is the integration time. Self-noise appears across the dirty map via sidelobes. The dirty map is a convolution of the true radio brightness distribution of the source $\I(\theta_x,\theta_y)$ with the point spread function (PSF), also called the ``dirty beam". Deconvolution of the PSF from the dirty map removes these sidelobes, yielding the so-called ``clean map" $\I_C^\circ(\theta_x,\theta_y)$, an estimate of the true radio brightness distribution. For a clean map the signal-to-noise ratio at a location on the source is given (Paper I) as SNR$ = I_C(\theta_x,\theta_y)\Omega_{\rm bm}/ \sigma_I(\theta_x,\theta_y)$ where $\Omega_{\rm bm}$ is the ``clean beam", usually well approximated by an elliptical Gaussian, and 

\begin{equation}
\sigma_I(\theta_x,\theta_y) = \frac{1}{M}\biggl[\I_C^\circ(\theta_x,\theta_y)\Omega_{\rm bm}+\frac{N}{n}\biggr]
\end{equation}
\noindent and the off-source \rms\ is simply $\sigma_I(\theta_x,\theta_y)=N/nM$. 

For reasons discussed in Paper~I, correlation arrays do not make total power measurements and the dirty map is instead given by:

\begin{equation}
\I_D(\theta_x,\theta_y)=\frac{1}{2n_b}\bigl\lbrace\sum_{j=1}^{n} \sum_{k>j}^{n}[\V_{jk}(u,v)e^{-i(\theta_x u+\theta_y v)}+\V_{jk}^\ast(u,v)e^{+i(\theta_x u+\theta_y v)}]\bigr\rbrace,
\end{equation}

\noindent where $n_b=n(n-1)/2$ is the number of unique antenna pairs. In this case, there is no simple expression for the map variance. \citet{Kulkarni1989} derived the map variance for the general case of an array with an arbitrary number of antennas $n$, and arbitrary values of $S$ and $N$. The results are expressed in terms of $[n_b\times n_b]$ covariance matrices which must be calculated explicitly for the array and source in question. Paper~I considered several simple cases, leading to the conclusion that for a point source observed by a correlation array, the dirty map $\I_D$ is the same as the scaled PSF. Its sidelobes and the associated self-noise can be removed from off-source regions through deconvolution. The on-source noise is then $\sigma_I^{on}=S/M$ and the off-source noise is $\sigma_I^{\rm off}=N/\sqrt{2n_b}M\approx N/nM$, similar to the result for a total power array. Unlike a total power array, however, if a point source is observed against strong uniform background emission by a correlation array, we have $\sigma_I^{\rm off}=(S+N)/\sqrt{2n_b}M$. The sidelobe response of the point source and its associated self-noise can again be removed from the map through deconvolution, but the noise floor remains. 

For an extended source the map \rms\ approaches 

\begin{equation}
\sigma_I(\theta_x,\theta_y) \approx \frac{1}{M}\biggl[\I_D(\theta_x,\theta_y)+\frac{(S+N)}{\sqrt{2n_b}}\biggr]
\end{equation}

\noindent as $n$ becomes large. Since $\I_D$ is a convolution of the true radio brightness distribution with the PSF, self-noise is present across the map as a result of sidelobes. Unlike the case of a total power array, the self-noise contribution from sidelobes in a map of an extended source produced by a correlation array can only be removed approximately through deconvolution. In addition, a key difference between a total power array and a correlation array is that, in the former, only $N$ is uncorrelated between antennas whereas in the latter both $N$ and $S$ are uncorrelated between antennas, resulting in a noise floor $(S+N)/\sqrt{2n_b}M$ for a correlation array compared to $N/nM$ for a total power array. We then have

\begin{equation}
\sigma_I(\theta_x,\theta_y)\approx \frac{1}{M} \bigl[\I_C(\theta_x,\theta_y)\Omega_{\rm bm}+\frac{S+N}{\sqrt{2n_b}}\bigr].
\end{equation}

\noindent The signal-to-noise ratio for a location on a strong, extended source is then  
\begin{equation}
{\rm SNR} \approx M \frac{I_C{(\theta_x,\theta_y)\Omega_{\rm bm}}}{I_C(\theta_x,\theta_y)\Omega_{\rm bm}+(S+N)/\sqrt{2n_b}}
\end{equation}

\noindent For an array with antennas distributed over a circular footprint of diameter $d$, $\theta_a=\theta_b\approx \lambda/d$ and so $\Omega_{\rm bm}\approx(\lambda/d)^2$. We can express the map \rms\ in terms of $T_b$ and $T_{\rm ant}$ as:

\begin{eqnarray}
\sigma_T(\theta_x,\theta_y)&\approx& \frac{1}{M} \Bigl[T_b(\theta_x,\theta_y)+\frac{T_{\rm ant}+T_{\rm sys}}{\sqrt{2n_b}} \frac{\lambda^2}{A_e}\frac{1}{\Omega_{\rm bm}}\Bigr]\nonumber\\
&\approx&\frac{1}{M} \Bigl[T_b(\theta_x,\theta_y)+\frac{T_{\rm ant}+T_{\rm sys}}{(nA_e/d^2)} \Bigr]
\end{eqnarray}

\noindent where the last expression is approximately valid as $n$ becomes large. We can regard $nA_e$ as being the total collecting area of the array while $d^2$ is the area covered by the array — its ``footprint''. Their ratio is then an array areal filling factor $f_a=nA_e/d^2\approx n\Omega_{\rm bm}/\Omega_{\rm FOV}$, where $\Omega_{\rm FOV}\sim  \lambda^2/A_e$ is the field of view of a single antenna. The SNR in terms of $T_b$, $T_{\rm ant}$, and $T_{\rm sys}$ is then 

\begin{equation}
SNR \approx M \frac{T_b(\theta_x,\theta_y)}{T_b(\theta_x,\theta_y)+(T_{\rm ant}+T_{\rm sys})/f_a}.
\end{equation}

\noindent From Equations~7 and 9 we see that it is always the case that the on-source SNR $ \le M$. 

For those locations in the map where  $\I_C\Omega_{\rm bm}<<(S+N)/\sqrt{2n_b}$ or $T_b << (T_{\rm ant}+T_{\rm sys})/f_a$ we have, for strong sources,  $\sigma_I\approx S/nM$ or $\sigma_T\approx T_{\rm ant}/f_aM$. If this condition is met everywhere in the map the \rms\ is essentially uniform across the map and the observation is analogous to one made by antennas with ``hot receivers". Note that, all other things being equal, as the footprint of the array increases $\Omega_{\rm bm}$ decreases and this condition will be met. The SNR in these cases is

\begin{equation}
{\rm SNR} \approx Mn \frac{\I_C(\theta_x,\theta_y)\Omega_{\rm bm}}{S} \approx M f_a \frac{T_b(\theta_x,\theta_y)}{T_{\rm ant}}.
\end{equation}

\noindent The dynamic range of a radio map is conventionally given as the ratio of the maximum signal to the off-source {\rms}; that is, the maximum SNR. As is implicit in the above discussion, and as we discuss further in later sections, this simple definition does not apply to sources dominated by self-noise. We must distinguish between the on-source dynamic range, which is always $M$ or less, and the more conventional sense above. The dynamic range is given by substituting the maximum flux per beam or the maximum observed brightness temperature $T_b^{\rm max}$ into the relevant expression above.

\section{Representative Telescopes}

Our focus is on radio interferometric arrays that are capable of observing the Sun at dm$-\lambda$ and cm$-\lambda$ although the framework presented for evaluating the impact of self-noise outlined in Paper~I and here can be applied to a radio telescope operating at any wavelength. In particular, we describe two telescopes active in the United States that have been used for solar observations for many years: the {Expanded Owens Valley Solar Array} (EOVSA) and the {Jansky Very Large Array} (JVLA). We also discuss two instruments that are in the planning stages, the {\sl Frequency Agile Solar Radiotelescope} (FASR) and the {next-generation Very Large Array} (ngVLA).  EOVSA and FASR are solar-dedicated and therefore differ in significant ways from the JVLA and the ngVLA, which are both general purpose telescopes but are, or will be, capable of observing the Sun. Tables~1 and 2 summarize key system parameters, including antenna diameter, number antennas $n$, the frequency range used to observe the Sun, the FOV at a reference frequency of 1 GHz, and the angular resolution of the array — again at a reference frequency of 1~GHz. Also given is the array filling factor $f_a$, computed under the assumption the antenna efficiency is 0.65 in all cases (see text). Note that the sensitivity parameter $N$ is given in SFU for EOVSA and FASR A, and in Jy for the JVLA and the ngVLA. 

\subsection{Solar Dedicated Telescopes}

\begin{table}
\caption{Critical Telescope Parameters.}
\begin{tabular}{lccccccc}
& N & $n$ & $D$ & Band & $\theta_{\rm FOV}$ & $\theta_{bm}$ & $f_a$ \\Instrument & (SFU) & & (m) & (\rm GHz) & (deg/$\nu_{\rm GHz}$) & (asec/$\nu_{GHz}$) &$(10^{-5})$ \\
\hline
EOVSA &  $\sim\!125$& 13 & 2 & 1-18 &10.3&54& 1.8\\
FASR A& $\sim\!60$ & 130 & 2 & 2-20 &10.3&20& 3\\
\hline
\end{tabular}
\end{table}

{\sl\noindent Expanded Owens Valley Solar Array}: EOVSA is described by \citet{Gary2018}. It is a fixed array of $13\times 2$~m antennas with a maximum baseline $d_{\rm max}=1.2$~km. EOVSA observes the full disk of the Sun as a dedicated instrument from 1-18 GHz. It does so by rapidly sweeping across the frequency band, dwelling on a given spectral window for 19~ms, and then re-tuning to a different spectral window in 1~ms. It can sweep across the entire frequency band in 1~s or observe a fixed spectral window with a bandwidth of 375~MHz (of which 325~MHz is used) at high cadence. The frequency window is divided into 4096 channels and those corrupted by interference are flagged. The remaining channels are averaged to a typical bandwidth of 41~MHz. The antennas are relatively insensitive, a factor of several thousand times less sensitive than VLA antennas due to their small $A_e$ and large $T_{\rm sys}$ ($\approx 600$~K), yielding $N\approx 125$~SFU (Table~1). At present, EOVSA produces maps in Stokes I polarization. After completion of an upgrade in 2026, EOVSA will have 15 antennas and will be capable of producing maps in both Stokes I and V polarization. 
\smallskip

{\sl\noindent Frequency Agile Solar Radiotelescope}: FASR has been a long term priority for the solar physics community \citep{Bastian2019}. It is intended to fully exploit the diagnostic potential of imaging spectropolarimetry at radio wavelengths across a broad frequency range, nominally 0.2-20 GHz, using two subsystems - one operating from roughly 2-20~GHz (FASR A) and the other from 0.2-2~GHz (FASR B) using antennas of perhaps 2~m and 6~m diameter, respectively. It, too, will observe the full disk of the Sun. It is anticipated that FASR~A will have spectral resolution and time resolution similar to that of the JVLA and the ngVLA. FASR A will comprise a much larger number of antennas than EOVSA, $n\sim 130$ or more,  distributed over an area with $d_{\rm max}\approx 3$~km. It will observe in Stokes I and V. The value of $N$ for FASR is expected to be somewhat better than that of EOVSA due to better receiver performance and aperture efficiency but it is uncertain at this point. We adopt a value of $N=60$~SFU. For the purposes of discussion here, we only consider the FASR~A subsystem since it is closest in frequency coverage to EOVSA, the JVLA, and the lower three frequency bands of the ngVLA (1.2-20.5~GHz). 
\smallskip

\subsection{General Purpose Telescopes}

\begin{table}
\caption{Critical Telescope Parameters.}
\begin{tabular}{lccccccc}
& N & $n$ & $D$ & Band & $\theta_{\rm FOV}$ & $\theta_{bm}$& $f_a$\\
& (Jy) &  & (m) & (GHz) & (amin/$\nu_{\rm GHz}$) & (asec/$\nu_{\rm GHz}$)  & $(10^{-3})$\\
\hline
Jansky VLA & 250-420 & 27 & 25& 1-18 &45&20 & 1\\
ngVLA core& 390-420 & 114 & 18& 1.2-116 &62&20 & 2.1\\
\hline
\end{tabular}
\end{table}

General purpose telescopes are designed to address a science program of broad astrophysical interest and therefore support large user communities and a diverse range of observing modes. They are optimized for the study of faint celestial sources of radio emission and therefore emphasize sensitivity achieved by minimizing $N$. 
\smallskip

{\sl \noindent Jansky Very Large Array (JVLA)}: The JVLA \citep{Perley2011} is a sensitive, general purpose radio telescope located in New Mexico. It can observe the Sun in five frequency bands spanning 1-18 GHz. The JVLA comprises $27\times 25$~m antennas. The antennas are configured into one of four standard array configurations. The two most compact configurations, most commonly used to observe the Sun, have maximum antenna baselines of $d_{\rm max}=1$~km (D configuration) and 3~km (C configuration). Table~1 gives the field of view of a 25~m antenna and the angular resolution of the array in the C~configuration. For example, the field of view is $45'$ at a frequency of 1~GHz but is only $3'$ at 15~GHz. The JVLA typically observes the Sun over an instantaneous bandwidth of 1 or 2~GHz with 512 or 1024 frequency channels. It can observe solar radio emission with a time resolution as short as 10~ms. It usually observes in dual-polarization mode, enabling the formation of the Stokes I and V polarization parameters. The values of N given in Table~1 are representative of those spanning the 1-18 GHz frequency range. Note that they are given in Jy units rather than SFU. As a general purpose telescope, the JVLA is only available for solar and solar wind studies for a few percent of its observing time at most. 
\smallskip

{\sl\noindent next generation Very Large Array (ngVLA)}: The ngVLA is an ambitious, general purpose radio telescope that will be the most sensitive of its kind \citep{Murphy2018}. The reference design\footnote{The ngVLA reference design is described in three volumes at https://ngvla.nrao.edu/page/refdesign. Although well-advanced, the reference design is subject to change.} comprises $214\times 18$~m telescopes configured in a fixed array spanning baselines up to 1000~km. However, 114 antennas will be deployed in a core configuration with a diameter of $\approx\!3$~km, providing an angular resolution similar to the JVLA in its C configuration. While the specifics of solar observing modes are in the planning stages, it is anticipated that it will be be able to observe the Sun with a time resolution and spectral resolution similar to that of the JVLA. In addition, it will be capable observing over a much wider frequency range, from 1.2-116 GHz, in both Stokes I and V. The value for $N$ anticipated for the ngVLA are somewhat higher than those of the JVLA, largely due to the difference in antenna size. The primary differences between the JVLA and the ngVLA are in the number of antennas available for imaging and the larger frequency range accessible for study. As is the case for the JVLA, the ngVLA will only be available for solar studies for a small fraction of the available observing time. 
\smallskip

\section{Self-Noise in Synthesis Maps of Solar Radio Sources}\label{Solar}

A lower limit to flux density $S$ incident on a given antenna is determined by radio emission from the quiet Sun during solar minimum. The brightness temperature spectrum of the Sun was measured from 1-18~GHz by \citet{Zirin1991} during solar minimum and calibrated against the Moon. More recently, \citet{Shimojo2017} summarized decades of well-calibrated measurements of the total flux density from the Sun made by Toyokawa Observatory and the Nobeyama Solar Radio Observatory at 1, 2, 3.75, and 9.4~GHz.  Figure~1 shows the measurements reported by \citet{Zirin1991}, converted to flux density measurements by assuming the Sun is a uniform disk with an effective radius that decreases by 6\% from 1~GHz to 20~GHz.  Also shown are the values of the total flux density reported by \citet{Shimojo2017}, averaged over five solar minima, as well as values of the total flux averaged over six solar maxima. During solar maximum, active regions contribute significant incremental flux that varies slowly in time — historically referred to as the ``slowly varying component" —  as regions emerge, decay, and rotate onto and off of the solar disk. During the course of a solar cycle the spectrum of the total flux of the non-flaring Sun lies somewhere between the two limits shown in Figure~1. The difference becomes relatively small as the observing frequency exceeds 10~GHz.  

\begin{figure}
\centerline{\includegraphics[width=0.75\textwidth]{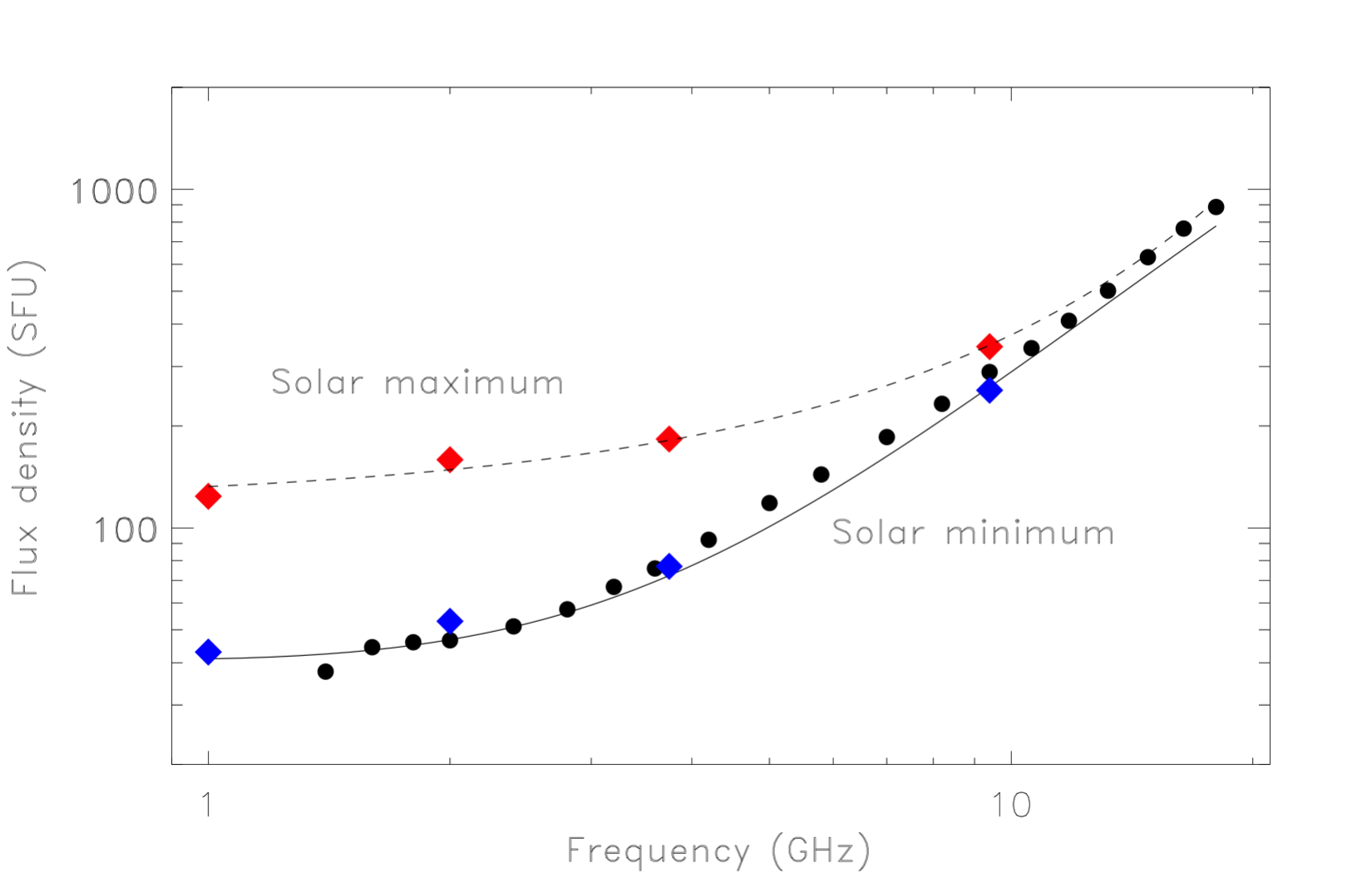}}
\caption{Spectrum of the total flux density from the quiet Sun. The {\it filled black circles} represent points measured by \citet{Zirin1991} during solar minimum, converted to a total flux density (see text). The {\it blue diamonds} show the total flux at fixed frequencies averaged over five solar minima; the {\it red diamonds} show the total flux at the same frequencies averaged of six solar maxima (see \citealt{Shimojo2017}).} 
\end{figure}

Figure~2a shows the solar flux $S$ as a function of frequency entering antennas from each of the instruments discussed in Section~2 for both solar minimum and solar maximum levels of activity. Figure~2b shows the corresponding antenna temperatures $T_{\rm ant}$. For EOVSA and FASR A, the 2~m antenna field of view is large enough to observe the full disk of the Sun across the entire 1-20 GHz frequency range, yielding similar values of $S$ for the two instruments. As is clear from Table 1, $S\sim N$ for both EOVSA and FASR A up to 10~GHz or so, and $S>N$ for higher frequencies.  It is always the case that $nS>>N$ for EOVSA and FASR A across the entire frequency range. In Paper~I it was found (see also \citealt{Kulkarni1989}) that self-noise becomes significant when $nS>N$. Note, however, that since $S\sim N$ over a significant range of frequencies for EOVSA, we must include $N$ or $T_{\rm sys}$ in estimates of the noise floor when considering EOVSA. 

In the case of the JVLA the constituent 25~m antennas resolve the Sun at frequencies $>\!1.5$~GHz. The single dish beam continues to decrease with frequency and so the source flux density $S$ entering the antenna decreases with frequency. Similarly, for the proposed ngVLA, the somewhat smaller 18~m antennas resolve the Sun at frequencies $\gtrsim2$~GHz. In both cases, $S$ decreases  from values comparable to those entering the 2~m antennas of EOVSA and FASR at 1~GHz, to $\sim\!10$ ~SFU at 20~GHz, a factor of $\approx 70$ smaller than is the case for EOVSA and FASR A. $N$ is very small for both the JVLA and the ngVLA by design (Table~2) and so $S>>N$ when observing the Sun.

We conclude that {\sl all} solar observations with current and planned arrays operating at \dml\ and \cml\ wavelengths, are dominated by self-noise to the extent that $nS>>N$ for EOVSA and FASR, and $S>>N$ for the JVLA and ngVLA, even for solar minimum conditions. Radio bursts and flares, of course, contribute much larger signals to the total. 

\begin{figure}
\centering
\includegraphics[width=0.485\textwidth]{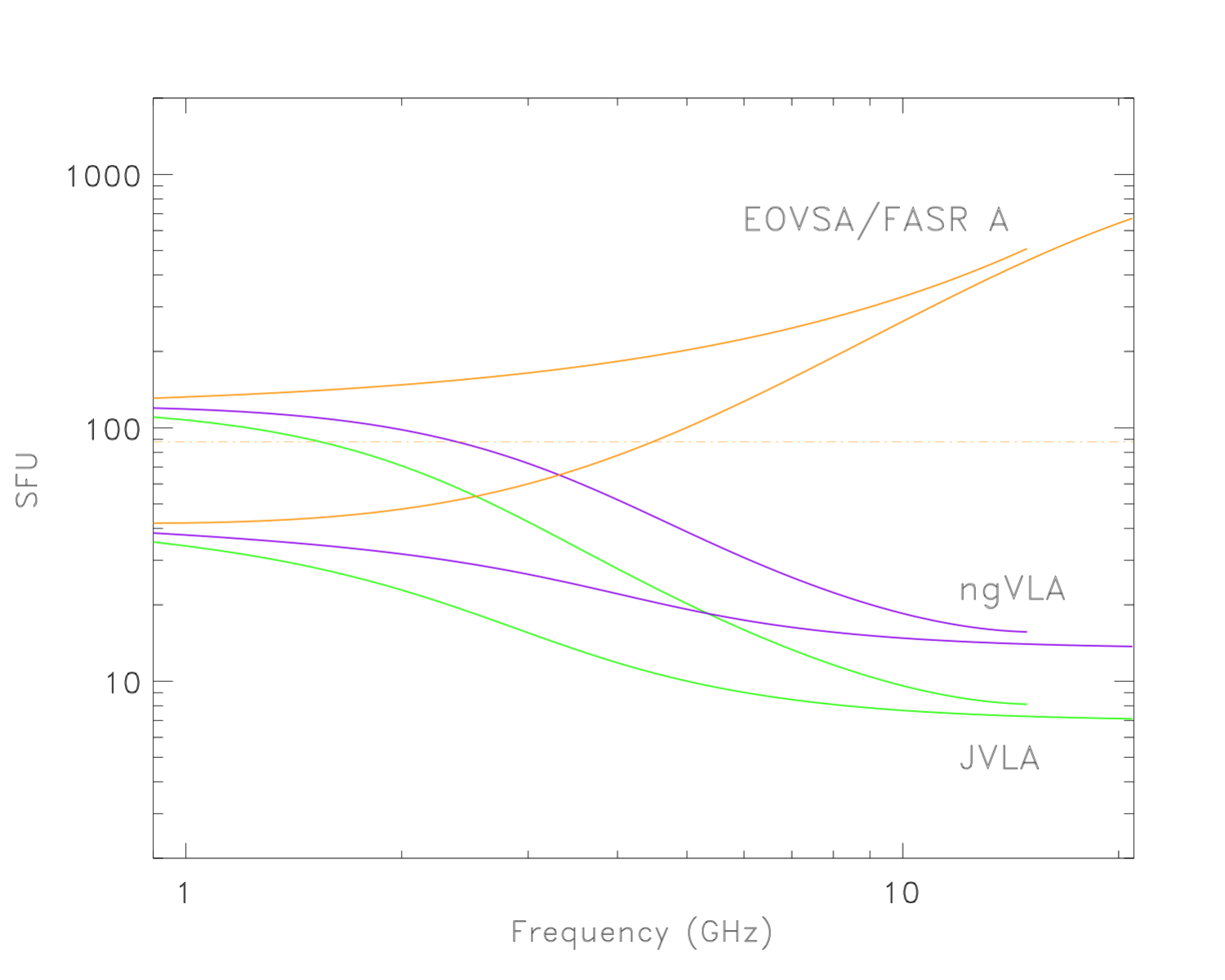}
\includegraphics[width=0.485\textwidth]{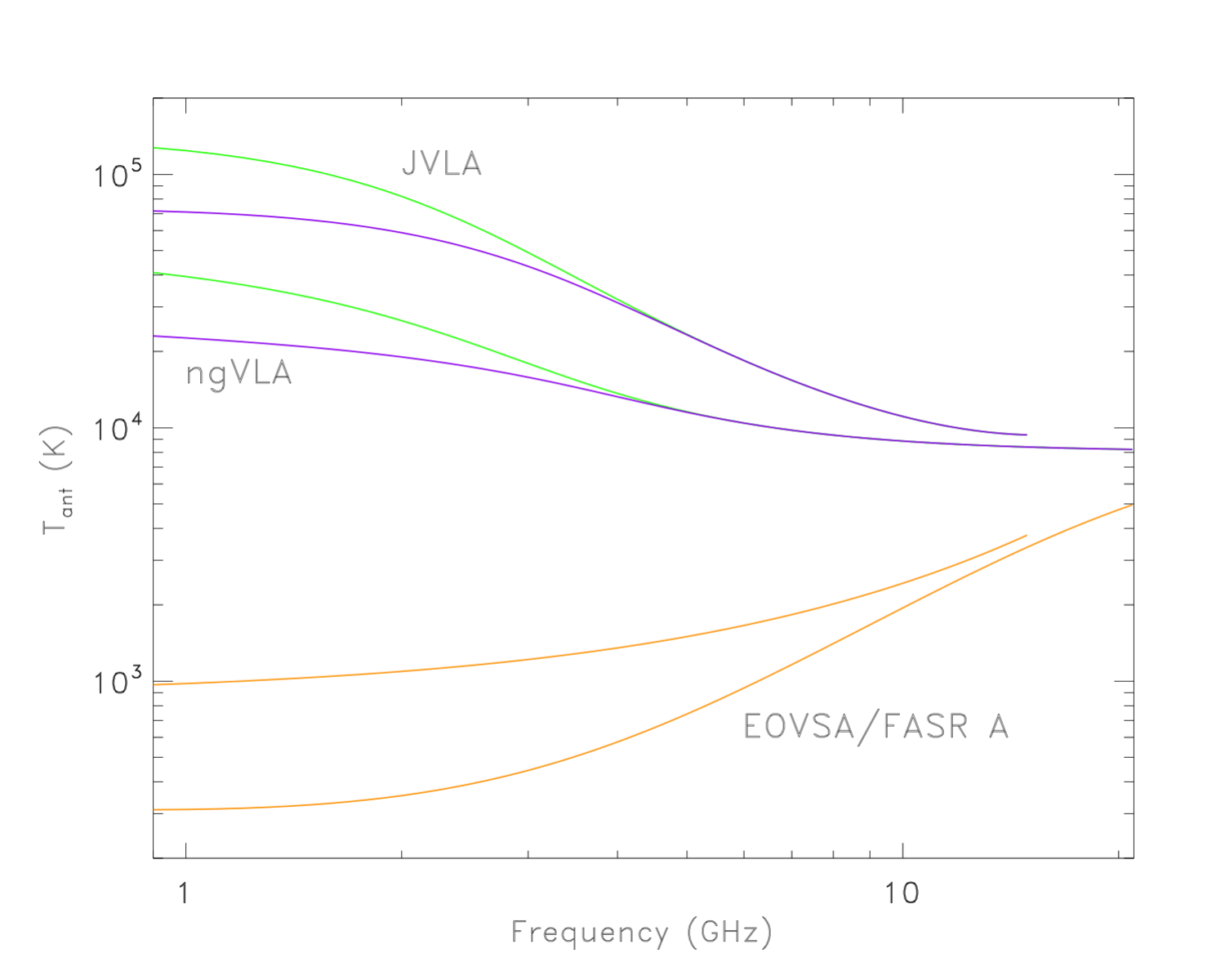}
\caption{a) The expected source signal $S$ as a function of frequency when observing the solar minimum and solar maximum spectra shown in Figure~2. For the JVLA and ngVLA, $N\sim 0.03$ SFU and is therefore not plotted. For EOVSA,  $N$ ({\it orange dashed line}) is comparable to $S$. This will also be true for an instrument like FASR. b) The corresponding values for the antenna temperature $T_{\rm ant}$. Notice that as the JVLA and ngVLA antennas resolve the Sun $T_{\rm ant}$ converges to the mean brightness temperature of the Sun.}
\end{figure}

\section{Science Use Cases} 

We now briefly consider several science use cases and consider the potential role of self-noise in each. We do so for two arrays currently available for such observations: EOVSA, which is solar dedicated, and the JVLA, which is a general purpose array.  We comment briefly on the two arrays in the planning stages, FASR (solar dedicated) and the ngVLA (general purpose) in Section~6. 

\subsection{Solar Radio Bursts}

Solar radio bursts have been studied since the earliest days of radio astronomy (see an early review by \citealt{Wild1963}). They tend to occur primarily at dm$-\lambda$ and longer wavelengths, or frequencies $\lesssim 3$~GHz. Examples of m-$\lambda$ bursts include those of type~II, driven by coronal shocks; and those of type~III, driven by nonthermal electron beams. Radio bursts at dm-$\lambda$ display a rich phenomenology (e.g., \citealt{Benz1991, Benz2001}). Some of these are analogs to their classical counterparts at m-$\lambda$ (e.g., type~IIIdm bursts) whereas it is less clear whether others have classical counterparts (e.g., the stochastic spike bursts reported by \citealt{Chen2015}). Instruments like LOFAR have provided new and exciting imaging observations of m-$\lambda$ radio bursts but dm-$\lambda$ imaging observations of bursts have been sparse. Radio bursts of all types are of interest because of their usefulness in diagnosing energetic phenomena on the Sun, including those with space weather impacts (see, for example, the review by \citealt{Gary2021}). Solar radio bursts are due to coherent emission mechanisms, either plasma radiation or cyclotron maser mechanisms \citep{Dulk1985, Bastian1998}. As such they can be very intense. An extreme example reported by \citet{Gary2021} produced a peak flux density of $10^6$~SFU, or $10^{10}$ Jy! Coherent radio sources are intrinsically unresolved. However, the observed source sizes at dm-$\lambda$ are significant due to scattering on electron density inhomogeneities in the corona \citep{Bastian1994}. Typical source sizes of $20-40"$ at dm$-\lambda$ are observed (e.g., \citealt{Chen2015}). While coherent bursts with flux densities as high as $10^6$ SFU are rare, radio bursts can often be 1000s to several $\times 10^4$~SFU. 

Dynamic spectroscopic observations show that coherent radio bursts at dm-$\lambda$ can be highly structured in the time-frequency domain. Snapshot integration times of $\sim\!10$~ms and spectral channel bandwidths of order 1~MHz are desirable to resolve such structure. Adopting these values for illustrative purposes, $M=100$. Let us suppose that a coherent radio burst occurs with a large flux density $S=10^4$~SFU at a wavelength of 20~cm (1.5~GHz) and that it is scatter-broadened into a Gaussian with a FWHM of $\theta_G=30"$. Since $n$ is not large for either EOVSA or the JVLA, we can compare approximate results with those resulting from the simplified expressions of \citet{Kulkarni1989} given in Paper I (Appendix B).

If observed by the JVLA, the antenna temperature would be $T_{\rm ant}\approx 10^7$~K and so $T_{\rm ant}/f_a\approx 10^{10}$~K. The angular resolution of the JVLA at a wavelength of 20~cm in the C configuration is $\theta_{\rm INT}\approx 13.3"$. We find that the on-source maximum is 2660~SFU per beam, corresponding to a brightness temperature of $T_b=8.7\times 10^{10}$~K, a lower limit because the source is scatter-broadened into a source that is larger than its intrinsic size. In this case $T_b>>T_{\rm ant}/f_a$ and so Equation~10 cannot be used to estimate SNR at the location of the burst. Using Equation~6, we estimate SNR $\approx 88$. Calculating map \rms\ exactly (Paper I, Appendix B), we find that $\sigma_I\approx 29$~SFU at the source maximum yielding an on-source SNR $\I_C(\theta_x)\Omega_{\rm bm}/\sigma_I = 92$, about 5\% larger than is given by Equation~6. Well off the source the clean map \rms\ is $\sigma_I\approx 3.2$~SFU and the formal dynamic range is DR$ =830$ although residual self-noise in the sidelobes remains. We note that the noise floor for a well-resolved source would be $\sigma_I \approx 10^4/M\sqrt{2n_b}=3.8$~SFU, yielding a somewhat lower DR$=720$ (Equation~10), and so the approximate expression the SNR is incorrect by $\lesssim 15\%$ for a strong source that is poorly resolved. 

The same burst observed by EOVSA with the same integration time and bandwidth would observe a peak of 3670~SFU per beam as a consequence of having a larger synthesized beam width of $\theta_{\rm INT}\approx 36"$. The corresponding brightness temperature is $1.9\times 10^{10}$~K, lower than the peak brightness temperature observed by the JVLA because it is barely resolved by EOVSA (beam dilution). The antenna temperature is $T_{ant}\approx 7.4\times 10^4$~K and $T_{ant}/f_a\approx 4\times 10^9$~K. We again compute the map \rms\ approximately using Equation~6 and find that $\sigma_I\approx 40.5$~ SFU and so the on-source is SNR$ =82 \lesssim\! M$, whereas the exact calculation yields SNR $=91$.  The approximate expression is in error by about 10\%. For positions far from the source, $\sigma_I\approx 5.4$~SFU and so the formal dynamic range in this case is $DR=680$, compared with the less accurate value of 460 using the expressions for a well-resolved source (Equation~10), a difference of about 30\%. The errors resulting from the use of the approximate expressions for SNR are greater for EOVSA than is the case for the JVLA as a consequence of EOVSA having fewer antennas.  


For this example, self-noise imposes an on-source limit of SNR$\lesssim M$ for both the JVLA and EOVSA and so the source flux can only be known to a few percent. The SNR is consistent with results from Paper I, that the on-source SNR is independent of the number and size of antennas in the array in this case. The formal dynamic range is of order 830 (680) for the JVLA (EOVSA). Features of a few times the off-source map \rms\ should be detectable for snapshot maps of the burst source that are perfectly deconvolved. For example, features in the field of view with $5\sigma$ brightness temperatures of  $5\times 10^8$~K ($1.4\times 10^8$~K) should be detectable by the JVLA (EOVSA). To image fainter emissions in the presence of radio bursts may require alternate strategies, as we discuss briefly in Section~6.


\subsection{Solar Flares}

Solar flares accelerate electrons to high energies, producing incoherent gyrosynchrotron (GS) radiation at dm- and cm-$\lambda$ \citep{Bastian1998}. Radio observations at these wavelengths are important for constraining electron acceleration and transport mechanisms as well as dynamic measurements of coronal magnetic fields in flaring sources (e.g., \citealt{Chen2020a}). A critical science use case for any next-generation solar radio telescope is to produce high-SNR and high-DR snapshot images of flare emissions across the relevant frequency range in order to enable these diagnostics with high precision \citep{Gary2023}. Flares can produce flux densities range from 10s of SFU to several times $10^4$ SFU \citep{Nita2002} and peak brightness temperatures of a few $\times\ 10^7-10^9$~K. While flare sources at cm-$\lambda$ are often relatively compact -- a few arcsec to 10s of arcsec -- they can be quite extended at dm-$\lambda$ \citep{Gary2018, Fleishman2018, Chen2020}. 

For illustrative purposes, we consider a ``flare'' that emits a total flux of 1000~SFU at a wavelength of 5~cm (6 GHz) from a source that we model as a core-halo structure.  Let the compact core be a Gaussian with a $\theta_G=10"$ and the halo component a Gaussian with $\theta_G=60"$. For simplicity the maxima of the two Gaussians coincide. We let the fractional flux density in the core increase from 0\%, 2\%, 5\%, 10\%, 20\%, to 50\% of the total flux; the diffuse component correspondingly decreases from $100\%$ of the total to $50\%$. We label these schematic models ``Source 1" through ``Source 6'' in Figs.~3 and 4. A GS spectrum does not typically show substructure on small spectral and temporal scales. We therefore relax the snapshot integration time to $\tau=1$~s and the bandwidth to $\Delta\nu=25$~MHz so that $M=5000$. We note in passing that EOVSA currently uses an integration time of 19~ms and an effective bandwidth of 41~MHz (Section~3.1) and so $M=880$ in practice. For the purposes of comparison, however, we use $M=5000$ for both instruments. 

In the case of the JVLA, the antenna temperature is $T_{\rm ant}=1.2\times 10^6$~K and $T_{\rm ant}/f_a= 1.2\times 10^9$. We find that the maximum map values range from 6.3 SFU/bm (Source 1), corresponding to a brightness temperature of $T_b\approx 2\times 10^8$~K to 100 SFU/bm (Source 6), corresponding to $T_b\approx 3.25\times 10^9$~K. When the diffuse component dominates $f_a T_b < T_{\rm ant}$ and so SNR $\approx Mf_a T_b/T_{\rm ant} \approx 420$ (Equation~10). In contrast, as the core component increases Equation~8 must be used, giving SNR $\approx M\ [3.25\times 10^9/(3.25\times 10^9+1.2\times 10^9)]=3650$. 

For EOVSA, the antenna temperature is $T_{\rm ant}=7400$~K and $T_{\rm ant}/f_a=4.1\times 10^8~K$. The peak flux density for Sources 1-6 ranges from 109 to 446~SFU/bm corresponding $T_b\approx 5.7\times 10^8$~K to $2.3\times 10^9$~K. For EOVSA, then, we are in a regime where Equation~8 is relevant for all sources.  The corresponding SNR ranges from 2900 for Source~1 to 4200 for Source~6. 

\begin{figure}
\centering
\includegraphics[width=0.9\textwidth]{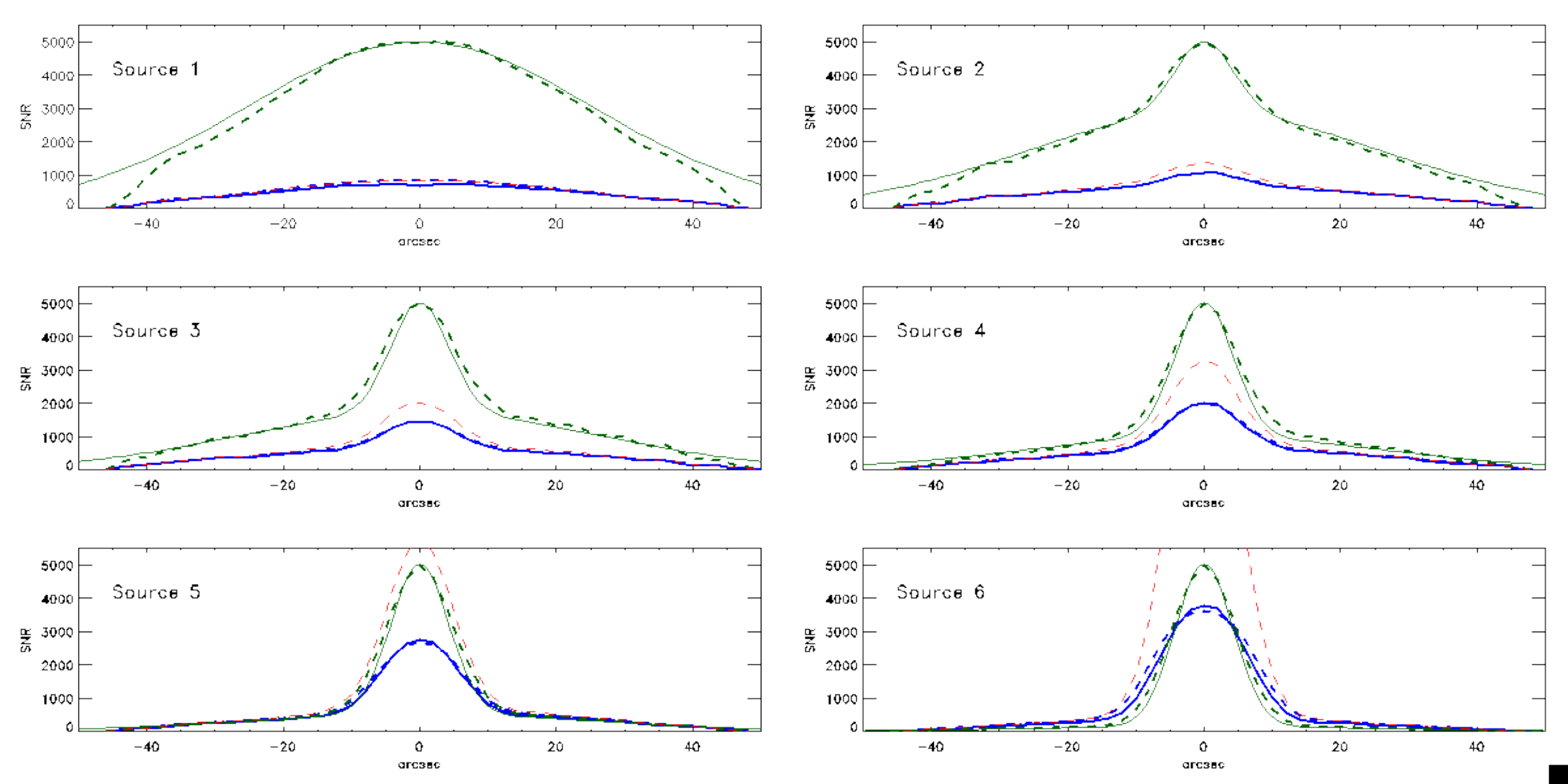}
\caption{The SNR expected for JVLA observations of a flare. The source is modeled as a ``core-halo" structure comprising two Gaussians with $\theta_G=60"$ (halo) and $\theta_G=10"$ (core). The fraction of the total emission attributed to the core is 0\%, 2\%, 5\%, 10\%, 20\%, and 50\% for sources 1-6, respectively. The {\it green line} represents the model and the {\it dashed green line} represents the clean map $\I_C\Omega_{\rm bm}$ (both in arbitrary units); the {\it blue line} represents the SNR calculated explicitly; the {\it dashed blue line} represents the approximate SNR calculated using Equation~7; and the {\it dashed red line} represents the SNR calculated using Equation~10.}
\end{figure}

These points are illustrated in Figs.~3 and 4. In all panels, the ordinate is scaled to SNR with a range of $M$. The solid green line shows the source model and the dashed green line shows the clean map $\I_C(\theta_x)\Omega_{\rm bm}$ of the core-halo model (arbitrary units); the solid blue line shows SNR computed from $\I_C(\theta_x)\Omega_{\rm bm}/\sigma_I$ where $\sigma_I$ has been computed using the exact expressions give in Paper I, Appendix B; the dashed blue line shows the SNR computed using Equation~7; and the dashed red line is the SNR computed from Equation~10.  For the JVLA, the fidelity of the clean maps is quite good although they deviate from the model at the edges of the map as the array over-resolves the source somewhat. The SNR is well-approximated by Equation~6 in all cases although it begins to deviate from the exact calculation by a few percent as the core component increases (Source 6).  Equation~10 (red dashed line) is a poor approximation of the SNR except for those cases dominated by the halo component. 

In the case of EOVSA, with the possible exception of Source~1, the fidelity of the clean map is poor: the dashed green lines indicate a much broader source than is actually the case (solid green line). For Sources 1-4 the SNR computed using Equation~7 or 9 (dashed blue lines) is similar to that computed explicitly (solid blue lines), but is systematically higher than the exact calculation by 35\% (Source~1) to 20\% (Source 4). As was the case for the JVLA, the SNR increases toward $M$ as the importance of the core component increases in Sources 5 and 6. Interestingly, the SNR is greater on each source as observed by EOVSA compared with the JVLA. This may be understood as the result of the EOVSA beam $\Omega_{\rm bm}$ being roughly a factor 7 larger than that of the JVLA as shown at the end of Section~2. 

\begin{figure}
\centering
\includegraphics[width=0.9\textwidth]{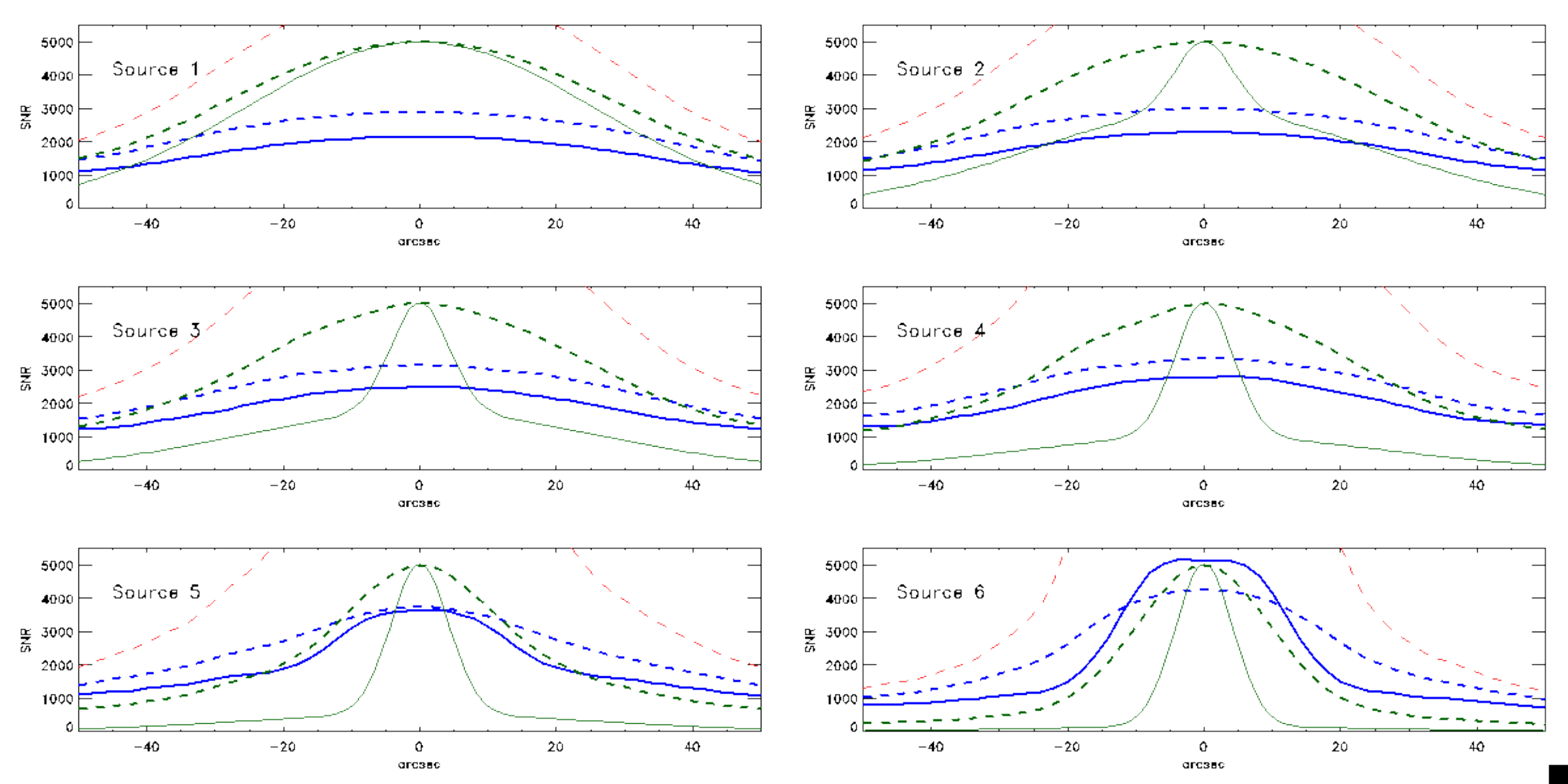}
\caption{The SNR expected for EOVSA observations of a flare. The source models are identical to those used in Figure~3, as are the various lines. }
\end{figure}

Self-noise present in snapshot dirty maps of extended sources through source sidelobes can be mitigated through deconvolution. For total power arrays it can be removed entirely; for correlation arrays it be removed to an increasing degree as the number of antennas $n$ in the array increases. We illustrate this in Figure~5 for the schematic flare sources. We estimate the residual noise in the clean maps of the flare sources as $(\sigma_I-\I_D)/M$ and compare it to the noise floor given by $S/M\sqrt{2n_b}\approx S/Mn$ or, equivalently, $T_{ant}/Mf_a$ (dashed red line). For the JVLA, the residual is close to the noise floor for the diffuse model (Source 1) but it shows increased sidelobe structure and lies somewhat below the noise floor for the Source 6. Nevertheless, for the JVLA the off-source \rms\ is well-approximated by the noise floor: $\sigma_I\approx S/\sqrt{2n_b}M=7.5\times 10^{-3}$~SFU ($2.4\times 10^5$~K). The dynamic range therefore formally ranges from DR$\approx 830$ to 13,000 for Sources~1-6. 

In the case of EOVSA, the residual deviates from the noise floor in all cases. The noise floor is $0.016$~SFU ($8.4\times 10^4$~K) but the calculated value for the off-source \rms\ ranges from roughly 0.5-0.75 this value. For EOVSA, then, the dynamic range formally ranges from DR$\approx 9000$ to 56,000 for sources 1-6. It may seem paradoxical that EOVSA has the higher DR. We expect, in general, that flares lie entirely within the field of view of both the JVLA and EOVSA from 1-18 GHz and so both 2~m and 25~m antennas intercept $S$. As was the case for SNR the ratio of the EOVSA DR to that of the JVLA is dominated by ratio of the synthesized beams; the difference is therefore primarily due to the difference in resolution although the details of the distribution of brightness also necessarily plays a role. 

The dynamic range is often used as a rough metric for the ratio between the maximum and minimum believable signal in a given map. When observing strong, extended  sources, however, the on-source SNR is limited to an upper limit of $M$. When evaluating the significance of emission features it is advisable to use the on-source SNR, not the DR, to assess the significance of emitting features. For relatively faint on-source emission we have $\sigma_T\approx T_{\rm ant}/Mf_a$.  For the above examples, $S=1000$~SFU, we would have $\sigma_T=1.7\times 10^5$ and $5\times 10^4$ for the JVLA and EOVSA, respectively. Sources with brightness temperatures of a few times $\sigma_T$ can be imaged in the presence of the flare in snapshot images. It is also worth emphasizing, however,  that the SNR and DR metrics can be quite misleading if the image fidelity is poor, as can be the case for snapshot images of complex sources made with sparse arrays.  

\begin{figure}
\centering
\includegraphics[width=0.45\textwidth]{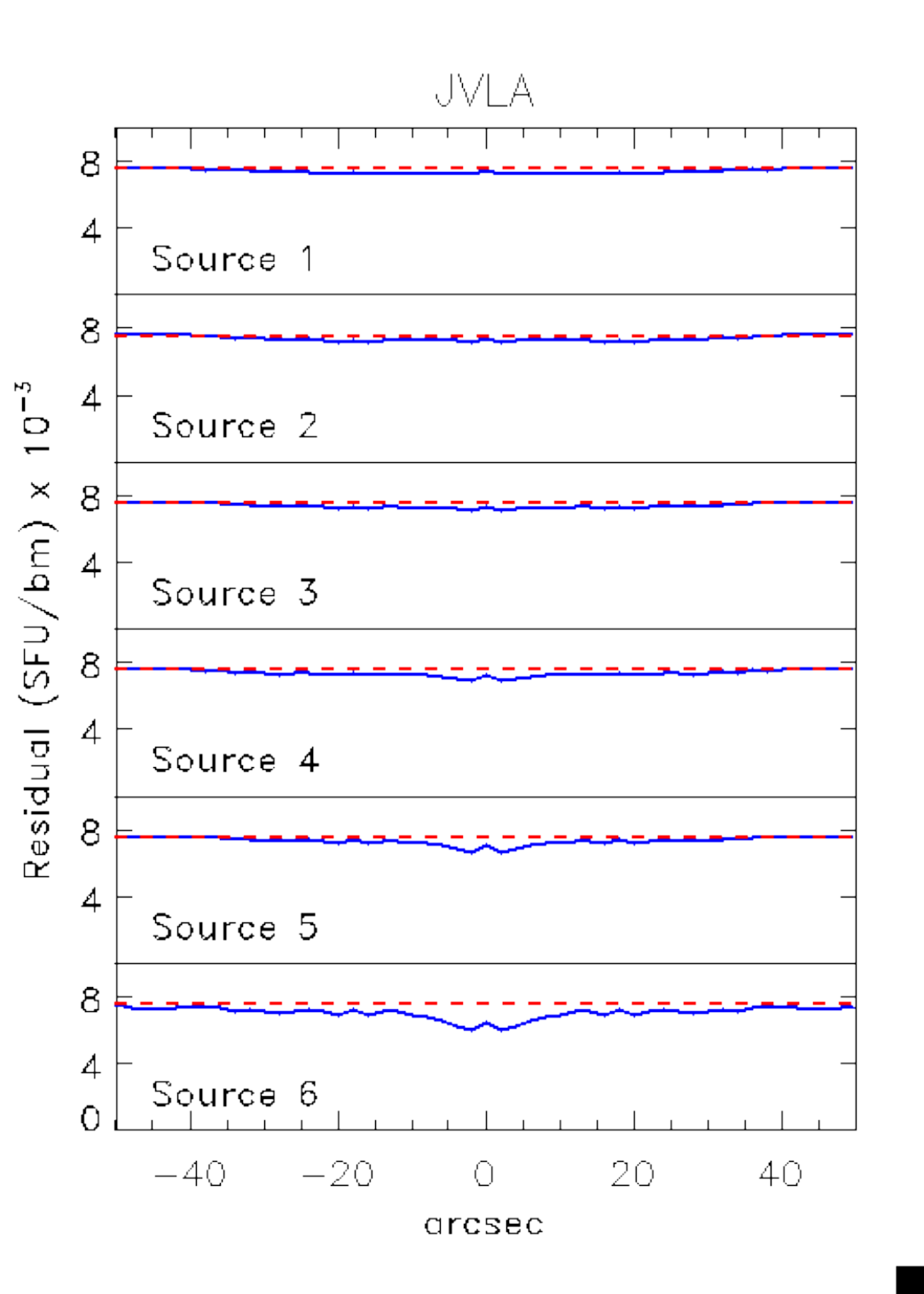}
\includegraphics[width=0.45\textwidth]{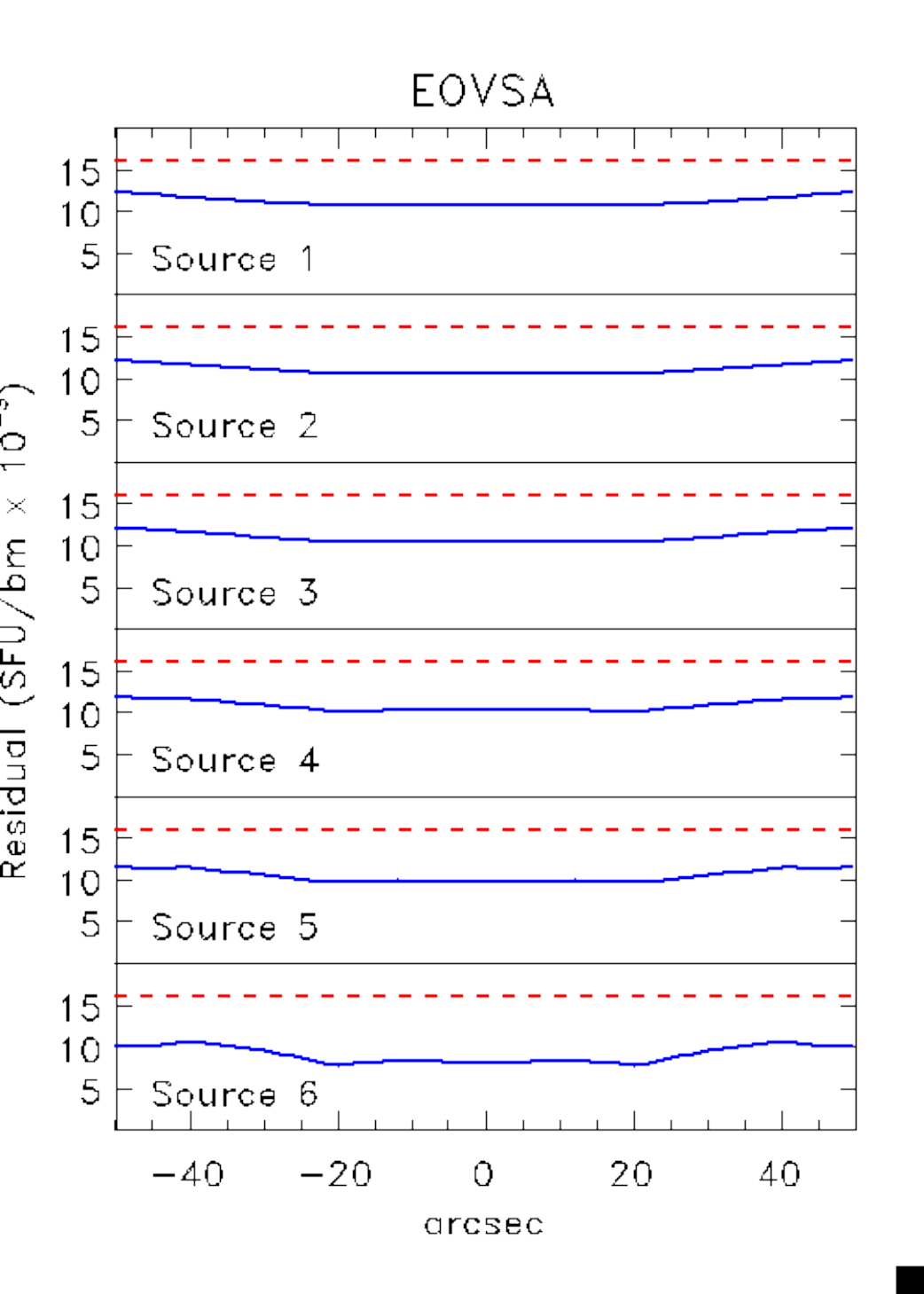}
\caption{Left: Comparison of the residual noise ({\it blue line}) in the clean map for each model source observed by the JVLA compared with the noise floor computed as $S/\sqrt{2n_b}M$ ({\it red dashed line}). Right: The same when the sources are observed by EOVSA. }

\end{figure}
\subsection{Microflares and Other Small Transients}

Small radio transients — e.g., microflares, jets  — have been observed against the background quiet Sun for many years (e.g., \citealt{Bastian1991, Gary1997, Kundu1997, Qiu2004, Sharma2020}). They occur on time scales similar to those relevant to impulsive flares. For illustrative purposes, therefore, we again let $\tau=1$ s and $\Delta\nu=25$~MHz so that $M=5000$.  They have flux densities ranging from $\sim 0.01$~SFU or less up to a few SFU; that is, they do not contribute significantly to $S$ or $T_{\rm ant}$. 

When observing the quiet Sun it is important to note that the 25~m JVLA antennas resolve the Sun to an increasing degree as the observing frequency exceeds 1.5~GHz while the Sun remains essentially unresolved by the EOVSA 2~m antennas across the 1-18 GHz frequency range. As a result, the quiet Sun flux density $S$ entering a JVLA antenna decreases with frequency while that entering an EOVSA antenna increases with frequency as shown in Figure~2a. The corresponding JVLA antenna temperature converges to the mean brightness temperature across the field of view while the EOVSA antenna temperature increases with frequency (Figure~2b). 

Consider a small transient with a flux density 0.01 SFU/beam observed at 6 GHz. Since, in this case, $\I_C\Omega_{\rm bm}<<S/n$ we can exploit Equation~9 to estimate the SNR. For the JVLA we have $S\approx 10$~SFU. The on-source SNR is well-approximated by $\sqrt{2n_b}M \I_C\Omega_{\rm bm}/S$ and so the on-source SNR$ \approx 130$. For EOVSA, S$ \approx 120$~SFU and $N\approx 125$ and, therefore, SNR$ \approx 3$. The JVLA with its large antennas has a considerable advantage over EOVSA for detecting weak transients against the background Sun at frequencies above a few GHz. Even at 1~GHz, where the incident flux density $S$ is comparable for the two telescopes, there is a factor of 5 advantage in the JVLA sensitivity over EOVSA as a result of the larger number JVLA antennas and its negligible value of $N$.  

\subsection{The Quiet Sun}

By ``quiet Sun" emission we mean radio emission from the Sun in the absence of bursts, flares, or other transients that can significantly perturb the signal entering the system. The quiet Sun therefore includes emission from the background solar disk, active regions, coronal holes, and prominences and filaments.  Emission from the quiet Sun in the absence of active regions and transients is due almost entirely to thermal free-free emission. For the frequency range discussed here, the high-frequency brightness is dominated by emission from the chromosphere while coronal emission dominates at low frequencies. 

Radio emission from solar active regions is also dominated by thermal radiation. At low frequencies ($\nu\lesssim 2-3$ GHz)  thermal free-free absorption dominates and the active region can be optically thick at coronal heights, resulting in brightness temperatures of $2-3\times 10^6$ K on the scale of the active region. Free-free absorption in the corona declines rapidly with frequency, giving way to thermal gyroresonance absorption in regions where the magnetic field is sufficiently strong to render the corona optically thick at low harmonics of the electron gyrofrequency (e.g., \citealt{White1997}), again yielding coronal source brightness temperatures, but on scales more typical of sunspot umbrae and penumbrae. Thermal GR emission is relevant from $\sim 2-20$~GHz. At yet higher frequencies, $\nu\gtrsim20$~GHz, chromospheric thermal free-free emission dominates emission from active regions.

Since quiet Sun emission varies slowly in time compared to active phenomena, the constraints on snapshot imaging can be further relaxed in terms of integration time. With $\tau=25$~s, for example, and $\Delta\nu=25$~MHz, M$=25000$. In the absence of transient solar activity, we take the maximum brightness temperatures likely to be encountered to be $T_b^{\rm max}\sim 2\times 10^6$~K. In the case of the JVLA, $f_a\approx 10^{-3}$ and so $f_a T_b^{max}\approx 2000$~K. Comparing with Figure~2b we see that it is always the case that $T_{\rm ant}> 2000$ over the 1-20 GHz frequency range. Since $S>>N$ the map \rms\ can be approximated by $\sigma_T\approx T_{\rm ant}/Mf_a$ and the SNR is given by Equation~10, ranging from roughly 500 at low frequencies to $\sim 50$ at high frequencies. In the case of EOVSA, $f_a T_b^{\rm max}\approx 36$~K. Since $S\sim N$ for quiet Sun emission observed by EOVSA, we compare with $T_{\rm ant}+T_{\rm sys}$ where $T_{\rm sys}\approx 600$~K; in this case, $\sigma_T\approx (T_{\rm ant}+T_{sys})/Mf_a$ and here, too, we can use Equation~10 to estimate the SNR. We find that it ranges from as high as 1800 at low frequencies to just a few at high frequencies.

For concreteness, at a wavelength of 10~cm (2 GHz) $S=30-80$~SFU for the JVLA. The corresponding antenna temperature is $T_{\rm ant}\approx 35-92\times 10^3$~K and so the on-source SNR $\approx 3-7\times 10^{-4}\ T_b$. For an active region with a brightness temperature $T_b=2\times 10^6$~K, SNR$\sim 600-1400$. For EOVSA, all other things being equal, $T_{\rm ant}\approx 400-1000$~K and so SNR$=4.5-11\times 10^{-4}\ T_b \sim 900-2300$. At 18~GHz, however, the JVLA has $T_{\rm ant}\sim 10^4$~K whereas for EOVSA, $T_{ant}\approx 4000$~K. In this case, thermal gyroresonance emission is rare and chromospheric brightness is dominated by thermal free-free emission. The JVLA and EOVSA both yield similar SNRs of order 3-10. The SNR is far less than $M$ in all cases for snapshot images of the quiet Sun. While the SNR may be deemed adequate at lower frequencies, it is poor at higher frequencies. More importantly, however, snapshot  images of quiet Sun emission with sparse arrays like EOVSA and the JVLA are extremely poor in terms of image fidelity.

Happily, for quiet Sun imaging, we are in a regime where the map \rms\ is well approximated by the noise floor $(S+N)/\sqrt{2n_b}M$; the noise is essentially uniform and incoherent across the map, just as it is for a correlation array observing weak sources. This allows observers to exploit techniques that greatly improve aperture coverage ({\sl uv} coverage) using the technique of Earth rotation (ER) aperture synthesis and/or multi-frequency (MF) synthesis. 

\begin{figure}
\centerline{\includegraphics[width=\textwidth]{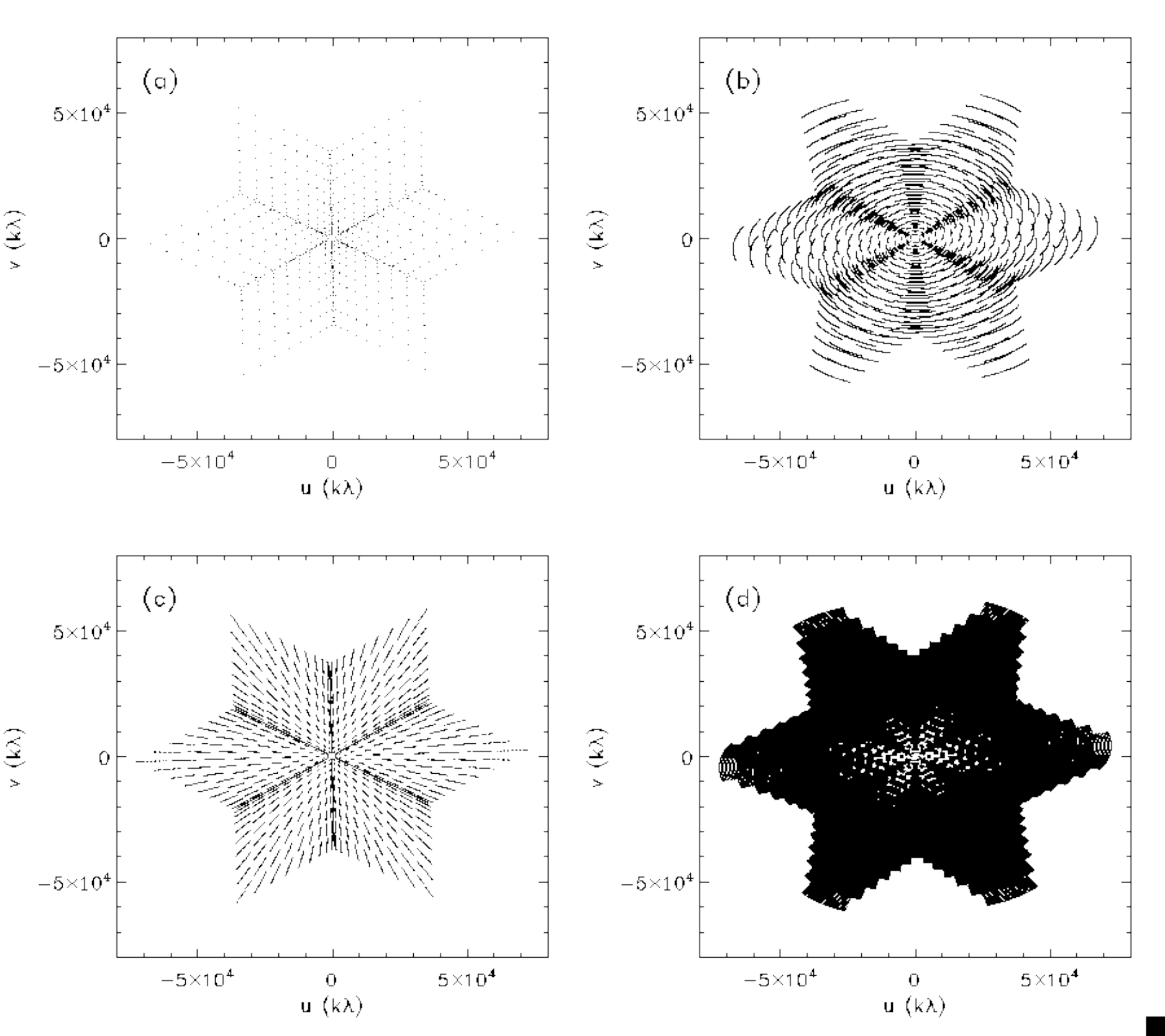}}
\caption{The {\sl uv} coverage provided by the JVLA in its C configuration. a) Snapshot coverage at the time of meridian passage at 6 GHz; b) ER synthesis coverage at 6 GHz for a period of 30 min; c) Snapshot coverage with MF synthesis from 5.5 to 6.5 GHz; a combination of ER and MF synthesis for 30 min and a frequency range of 5.5 to 6.5 GHz.}
\end{figure}

ER synthesis exploits the fact that the array geometry, as viewed from a celestial object, changes as the Earth rotates. Specifically, a given antenna baseline traces out a portion of an ellipse in the {\sl uv} plane with time as the Earth rotates (e.g., \citealt{Thompson1986}), thereby greatly improving sampling of the Fourier domain (or {\sl uv} plane). Most interferometric radio arrays exploit ER synthesis to improve sampling of the {\sl uv} plane which, in turn, improves imaging fidelity and dynamic range.  Implicit in the use of the technique is the assumption that noise is uncorrelated between baselines and that the radio brightness distribution of the sky does not change during the course of the synthesis; i.e., the source is static.

\citet{Vivek1991} discussed ER synthesis of strong sources where a significant correlated noise component is present in snapshot maps, thereby complicating its removal.  In the case of quiet Sun imaging, however, the self-noise in a snapshot map manifests as uncorrelated noise. This being the case, the noise present in each snapshot map sums incoherently in the ER synthesis map. If the integration time is $\tau$ and the quiet Sun is observed for a time $\Delta T$, a total of $m=\Delta T/\tau$ snapshots contribute to the map and the \rms\ noise is reduced by a factor $\sqrt{m}$. Hence, not only does ER synthesis improve the {\sl uv} coverage and image fidelity, it reduces the map \rms\ and hence increases the SNR and DR. For example, suppose the integration time of an observation remained at 25~s but an ER synthesis was performed for 2~hr. Then $m=288$ and the SNR would be improved by a factor $m^{1/2}\approx 17$, at least in principle. In practice, the longer the ER observation, the more likely that source variability compromises the SNR. 

MF synthesis (e.g., \citealt{Sault1999, Rau2011}) exploits the fact that for a fixed antenna baseline, variation of the observing frequency effectively changes the baseline length because it is measured in wavelengths. While ER synthesis improves {\sl uv} coverage in a quasi-azimuthal fashion, MF synthesis improves {\sl uv} coverage in the radial dimension. By observing the source with a given array at multiple frequencies the {\sl uv} coverage is increased in the radial dimension. A detailed discussion of ER and FR synthesis is lies outside the scope of this paper. Their use is largely limited to quiet Sun imaging and careful consideration must, in any case, be given not only the integration time $\tau$, and instantaneous frequency bandwidth $\Delta\nu$, but also the duration of the ER synthesis observation $T$ and the total frequency range sampled if MF synthesis is also used. Figure~6 shows the improvement in {\sl uv} coverage realized for 2~hr of ER synthesis, 1~GHz of MF synthesis, and the combination of both. 

\section{Summary and Discussion}

We have considered the properties of noise in snapshot maps of solar phenomena, for which the self noise must always be considered. The details of the distribution of noise depend on the radio brightness distribution in the field of view and the array used to observe it. Using the expressions of \citet{Kulkarni1989} it is possible to calculate the map \rms\ explicitly for a correlation array. This is entirely feasible for arrays comprising low to moderate numbers of antennas such as EOVSA ($n=13$) or the JVLA ($n=27$). However, as the number of antennas $n$ increases the number of elements in the covariance matrix increases as $n^4$ and the calculation rapidly becomes impractical for large-$n$ arrays. We find, however, that as $n$ increases the map \rms\ for extended sources approaches Equations~6 or 8, from which approximate expressions were derived for the SNR of cleaned snapshot maps (Equations~7 and 9). The DR can be expressed by Equation~10. A number of science use cases were considered for illustrative purposes from which we draw the following conclusions:

\begin{enumerate}
\item{\sl Solar radio bursts} are intense sources of radio emission from \dml\ to \ml. Their intrinsic source size is believed to be very small but they are scatter-broadened into apparent source sizes that can be resolved by modern arrays. In order to resolve radio bursts temporally and spectrally short integration times and narrow frequency bandwidths must be employed, resulting in relatively small values of $M$. We used $M=100$ for illustrative purposes. For somewhat resolved solar radio bursts we find that the on-source SNR$\sim M$, the maximum possible value. The DR is given approximately by Equation~10. 
\item{\sl Solar flares} were modeled schematically as a core-halo morphology for which fractional contribution of the core component ($\theta_G=10"$) to the total flux ($S=1000$~SFU) varied from 0\% to 50\% (Sources 1-6; Figs.~3 and 4). We found that, for the JVLA, the approximate values of the SNR were within a few percent of the exact SNR but they were less accurate for EOVSA. The on-source SNR was significantly less than $M$ when the halo component dominated the source but approached $M$ as the core component increased. Equation~10 could only be used to estimate the JVLA SNR when the core component was small. The off-source \rms\ is well approximated by $S/nM$ for the JVLA but is less accurate for EOVSA. The formal DR can be very high for flare sources.
\item{\sl Microflares and transients} represent a special case to the extent that the array antenna size comes into play. For the JVLA the Sun is resolved by its 25~m antennas for frequencies greater than 1.5~GHz whereas the Sun remains unresolved by EOVSA's 2~m antennas. As a consequence, the JVLA is significantly more sensitive to small transient sources than is EOVSA. 
\item{\sl The quiet Sun} is taken to be emission in the absence of active phenomena. It therefore includes emission from the quiet chromosphere, corona, coronal holes, quiescent filaments, and active regions. We find that Equation~10 is a good approximation for the SNR and DR. Since the Sun spans an angular size of order $0.5^\circ$ and can present a complex distribution of radio brightness due to the myriad features mentioned, relatively sparse arrays like EOVSA and the JVLA do a poor job imaging the quiet Sun in snapshot mode. The fact that the noise in snapshot maps of quiet Sun emission is uncorrelated means that ER or MF aperture can be exploited, with appropriate care, to greatly improve imaging fidelity and dynamic range by improving the sampling of the {\sl uv} plane. For small- to moderate-$n$ arrays like EOVSA and the JVLA, even ER and MF synthesis come with pitfalls. Relatively long observations are needed to produce higher fidelity maps and source variability may then increase the effective \rms. Similarly, spectral variation with location across a map may compromise the effectiveness of MF synthesis. 
\end{enumerate}

Next generation instrumentation was described in Section~3. The ngVLA is a general purpose radio telescope whereas FASR is intended to be solar dedicated. The main difference, for our purposes, between current and next generation instruments is the significant increase in the number of antennas in each array: the ngVLA  core array $n$ will increase by a factor of more than 4, and $n_b$ will increase by a factor of 18 compared to the JVLA.  For FASR A, the number of antennas is expected to be at least 10 times greater than EOVSA, with a concomitant increase in $n_b$ by a factor of 100 or more. Note that FASR A will also be a much higher resolution instrument than EOVSA, with an angular resolution comparable to the JVLA (C configuration) and the ngVLA core. The increase in $n$ and $n_b$ for both instruments will greatly improve the snapshot imaging fidelity of transient emissions. Nevertheless, the on-source SNR will always be less than $M$, no matter the number and size of antennas. Approaching this limit requires minimizing the contribution  of $(S+N)/M\sqrt{2n_b}\approx S/Mn$ to the map \rms. This can only be accomplished by increasing $n$ and maximizing $M$ within the constraints imposed by the science objectives of a particular use case. While ER/MF synthesis will still be needed for quiet Sun imaging by next-generation instruments, the time required to produce excellent images will be significantly shorter than is currently required by the JVLA or EOVSA. 

Looking forward, the framework provided here for evaluating self-noise can be used in developing the science requirements for instruments like FASR. Detailed simulations of snapshot images and ER/MF synthesis images of a variety of science use cases with model array configurations are needed to fully understand the opportunities available to exploit myriad radio diagnostics to study the Sun, to understand their limitations, and to develop observing strategies to maximize the scientific return of next generation instrumentation. 
 
\begin{acks}
The National Radio Astronomy Observatory is a facility of the National Science Foundation operated under cooperative agreement by Associated Universities, Inc. The Expanded Owens Valley Solar Array (EOVSA) was designed, built, and is now operated by the New Jersey Institute of Technology (NJIT) as a community facility. 
\end{acks}

\begin{fundinginformation}
EOVSA operations are supported by NSF grant AGS-2436999 and NASA grant 80NSSC20K0026 to NJIT.
\end{fundinginformation}

\bibliography{Noise.bib}{}
\bibliographystyle{aasjournal}

\end{article}
\end{document}